\documentclass{article}
\usepackage{graphicx}
\usepackage{amsmath}

\newtheorem{theorem}{Theorem}
\newtheorem{acknowledgement}[theorem]{Acknowledgement}

\newtheorem{proposition}[theorem]{Proposition}

\input{tcilatex}

\begin{document}

\title{Projective Formalism and some Methods from Algebraic Geometry in the Theory
of Gravitation}
\author{B. G. Dimitrov \\
\textit{Bogoliubov Laboratory for Theoretical Physics}\\
\textit{Joint Institute for Nuclear Research}\\
\textit{Dubna 141 980, Russia}\\
\textit{E-mail: bogdan@thsun1.jinr.ru}}
\maketitle

\begin{abstract}
The purpose of this paper is to propose the implementation of some methods
from algebraic geometry in the theory of gravitation, and more especially in
the variational formalism. It has been assumed that the metric tensor
depends on two vector fields, defined on a manifold, and also that the
gravitational Lagrangian depends on the metric tensor and its first and
second differentials (instead on the partial or covariant derivatives, as
usually assumed).

\ \ \ \ Assuming also different operators of variation and differentiation,
it has been shown that the first variation of the gravitational Lagrangian
can be represented as a third-rank polynomial in respect to the variables,
defined in terms of the variated or differentiated vector fields. Therefore,
the solution of the variational problem is found to be equivalent to finding
all the variables - elements of an algebraic variety, which satisfy the
algebraic equation.
\end{abstract}

\section{\protect\bigskip\ Introduction and statement of the investigated
problem.}

\noindent\ \ \ \ \ \ \ \ \ Variational approach is an essential and powerful
method [1, 2] in the theory of gravitation, not only for obtaining the
Einstein equations, but also the equations of motion in the case of some
space-time decomposition, for example the (3+1) splitting \ [3] of
space-time. Presently, another space-time splitting - (4+1) is frequently
used within the framework of the Kaluza-Klein theories, called
Randall-Sundrum models [4, 5], where the main idea is that the
four-dimensional Universe may have appeared as a result of a
compactification of a five-dimensional one with a line element, given by

\begin{equation}
ds^{2}=e^{-2kr_{c}r_{5}}\eta ^{\mu \nu }dx^{\mu }dx^{\nu
}+r_{c}^{2}dx_{5}^{2}\text{ ,}  \tag{1}
\end{equation}
where $r_{c}$ is a compactification radius, $\eta ^{\mu \nu }$ is the
ordinary Minkowski metric, $x_{5}\subset \left[ 0,\pi \right] $ is a
periodic coordinate and $\mu \nu $ are four dimensional indices. In both
cases, the gravitational Lagrangian is of the form
\begin{equation}
L\equiv -\sqrt{-g}R\equiv L(g_{ij},g_{ij;k},g_{ij;kl})\text{ ,}  \tag{2}
\end{equation}
where $;$ may denote either a partial or a covariant derivative as a result
of the standard scalar curvature representation.Let us decompose the metric
tensor $g_{ij}$ according to the known formulae

\begin{equation}
g_{ij}\equiv p_{ij}-\frac{1}{e}u_{i}u_{j}\text{ \ ,}  \tag{3}
\end{equation}
where $p_{ij}$ is the projective tensor (satisfying the projective relation $%
p_{i}^{k}p_{k}^{j}=p_{i}^{j}\neq \delta _{i}^{j})$, $u_{i}$ are the
(covariant) components of the vector field $u$ with a lenght $e=u_{i}u^{i}$.
If formulae (3) is applied in both cases,a substantial difference will be
noted. In the ADM\ (3+1) \ approach, the vector field components are
identified with some components of $g_{ij}$ so that $p_{i}^{j}=\delta
_{i}^{j}$ and the projective tensor components turn out to coincide with the
three-dimensional components $g_{ij}^{(3)}$, defined on the
three-dimensional submanifold, i.e. $p_{ij}=g_{ij}^{(3)}$. In a (4+1)
space-time splitting, or in some other kind of decomposition, for example,
warped compactifications [6] to four dimensional Minkowski space on
seven-dimensional manifolds, this will be of course no longer the case.The
already known \ nice geometrical meaning of the embedded space will not be
valid, and one will have to deal with some kind of a multidimensional
projective formalism and a decomposition

\begin{equation}
g_{ij}=p_{ij}+h_{ij}\text{ \ \ \ ,}  \tag{4}
\end{equation}
where space-time is decomposed into a $p$-dimensional subspace and
orthogonal to it $(n-p)-$dimensional space [7]. If we restrict ourselves
only to the gravitational part in the action (it is present in all mentioned
cases in its standard form), then the combined system of equations of motion
for $p_{ij\text{ }}$and for $u_{i}$ (or respectively, for $h_{ij}$) have to
be solved. Another example in the same spirit is from relativistic
hydrodynamics, where the vector field $u$ will be the tangent vector,
defined at each point of the trajectory of motion of matter.

However, there is also another ''alternative'' , and it shall be
investigated in the present paper. Namely, in view of the expression (3),
let us simply assume that $g_{ij}$ depends on two vector fields $%
u=u(x_{1},x_{2},......x_{n})$ and $v=v(x_{1},x_{2},....x_{n})$ :

\begin{equation}
g_{ij}(x_{1},x_{2},....x_{n})=g_{ij}(\overrightarrow{u},\overrightarrow{v})%
\text{ \ \ \ \ .}  \tag{5}
\end{equation}
The left-hand side suggests that $g_{ij}$ may be regarded (for each pair of $%
i$ and $j$) as some hypersurfaces on the $n-$dimensional manifold, but they
may also be unterpreted as defined on a two-dimensional manifold,
represented by the vector fields $u$ and $\ v$. In terms of the
differentials $du$ and $dv$, definite differential-geometric characteristics
may be assigned, such as the first and the second quadratic form [8]. This
is the reason why in the present paper the choice of the variables willl be
related with differential quantities.

\section{\protect\bigskip First and Second Differentials and Variations}

\bigskip In general, the differential of a vector is not necessarily a
vector. Here it shall be assumed that $du$ and $dv$ are defined in the
corresponding \textbf{tangent spaces}\ $T_{u}$ and $T_{v\text{ }}$ of the
vector fields $u$ and $v$. The \textbf{first differential} $dg_{ij}(u,v)$
will be given by the expression

\begin{equation}
dg_{ij}(u,v)=\frac{\partial g_{ij}}{\partial u^{k}}du^{k}+\frac{\partial
g_{ij}}{\partial v^{k}}dv^{k}=\frac{\partial g_{ij}}{\partial u}\mathbf{du}+%
\frac{\partial g_{ij}}{\partial v}\mathbf{dv}  \tag{6}
\end{equation}
(for brevity further the indices $k$ will be omitted). In the case when $%
\frac{\partial g_{ij}}{\partial u^{k}}(x_{1}^{0},...x_{n}^{0})$ and $\frac{%
\partial g_{ij}}{\partial v^{k}}(x_{1}^{0},....x_{n}^{0})$ \ \ \ \ form a
basis in the tangent space to the hypersurface $%
g_{ij}(x_{1},x_{2},....x_{n}) $, $\mathbf{du}$ and $\mathbf{dv}$ may also be
interpreted as the linear coordinates of the tangent vector. It can easily
be proved that in a curved space the differentials of a vector, having
vector transformation properties are only the \textbf{covariant differentials%
}. The kind of the differential is of no importance for the presented
formalism in this paper. Also, instead of $\mathbf{u}$ and $\mathbf{v}$ one
may write down some other (tensor) variable, related for example to a$k-$%
dimensional hypersurface, embedded in the $n-$dimensional spacetime.

Of more importance is the expression for the \textbf{second differential }

\begin{equation}
d^{2}g_{ij}(u,v)=\frac{\partial ^{2}g_{ij}}{\partial u^{2}}(\mathbf{du)}%
^{2}+2\frac{\partial g_{ij}}{\partial u}\mathbf{.}\frac{\partial g_{ij}}{%
\partial v}\mathbf{dudv+}\frac{\partial ^{2}g_{ij}}{\partial v^{2}}(\mathbf{%
dv)}^{2}\text{ \ \ \ .}  \tag{7}
\end{equation}
The expressions for the \textbf{first and second variations} of $g_{ij}(%
\mathbf{u,v})$ \ have the same structure. Note that $\frac{\partial
^{2}g_{ij}}{\partial u^{2}}(\mathbf{du)}^{2}$ is the concise notation for
\begin{equation}
\frac{\partial ^{2}g_{ij}}{\partial u^{2}}(\mathbf{du)}^{2}\equiv \frac{%
\partial ^{2}g_{ij}}{\partial u^{i}\partial u^{k}}du^{i}du^{k}\mathbf{+}%
\frac{\partial ^{2}g_{ij}}{\partial u^{i2}}\text{ }(du^{i})^{2}\text{,}
\tag{8}
\end{equation}
and also it has been assumed that $\delta ^{2}u\equiv d^{2}u\mathbf{\equiv }%
\delta ^{2}v\equiv d^{2}v\equiv 0$. If $du=a_{i}dx^{i}$, then we will have

\begin{equation}
d^{2}u=a_{i}d^{2}x^{i}+da_{i}\wedge dx^{i}=a_{i}d^{2}x^{i}+(\frac{\partial
a_{i}}{\partial x^{k}}-\frac{\partial a_{k}}{\partial x^{i}})dx^{i}dx^{k}=0%
\text{ \ \ \ .}  \tag{9}
\end{equation}
Provided the Poincare's theorem is fulfilled and $d^{2}x^{i}\equiv 0$ (i.e. $%
dx^{i}=const.-\mathbf{dx}$ is a \textbf{coordinate line}), one would have $%
d^{2}u=0$ only if

\begin{equation}
\frac{\partial a_{i}}{\partial x^{k}}-\frac{\partial a_{k}}{\partial x^{i}}%
\equiv 0  \tag{10}
\end{equation}
for every pair of $i$ and $k$. This means that $du$ is either an \textbf{%
exact differential} $(a_{i}=const.)$, or that the vector field $\mathbf{du}$
has \textbf{zero-helicity components }$a_{i}$ ($rot\mathbf{a\equiv 0}$,
eq.(10)) in regard to the chozen basic vector field components $dx^{i}$.
Note that if $rot\mathbf{a\equiv 0}$, but $dx^{i}$ \textbf{are not} basic
vectors, one would have $\mathbf{d}^{2}\mathbf{u\neq }0$. From a
mathematical point of view, models with $d^{2}x^{i}\neq 0,$ when the scalar
product $<e_{i},$ $dx^{j}>=f_{i}^{j}\neq \delta _{i}^{j}$ have been
considered in [9].

\section{\protect\bigskip Formulation of the Variational Problem in the Case
of Different Operators of Variation and Differentiation.}

The gravitational part of the action, which will be investigated, is of the
type

\begin{equation}
S\equiv \int L(g_{ij},dg_{ij},d^{2}g_{ij})d^{n}x\text{ \ \ \ \ \ \ \ \ .}
\tag{11}
\end{equation}
The \textbf{first variation of the action (}provided the volume element is
not being varied) is

\begin{equation}
\delta S\equiv \int \delta Ld^{n}x\equiv \int \left[ \frac{\partial L}{%
\partial g_{ij}}\delta g_{ij}+\frac{\partial L}{\partial (dg_{ij})}\delta
dg_{ij}+\frac{\partial L}{\partial (d^{2}g_{ij})}\delta d^{2}g_{ij}\right]
d^{n}x\text{ \ \ \ \ \ .}  \tag{12}
\end{equation}
The operators $\delta $ and $d$ are defined in one and the same way, but
here they are distinguished. In the spirit of Cartan's works [10], they may
correspond to variations and differentiations along different paths (not
necessaril $u$ and $v)$. For example, one of the operators may be defined on
a submanifold.

To find the equations of motion for $u$ and the divergent terms, one would
need to interchange the places of the operators $d$ and $\delta $. For that
purpose, the corresponding expressions for $\left[ \delta ,d\right] g_{ij}$
and $\left( \delta d^{2}-d\delta ^{2}\right) g_{ij}$ have to be found. The
\textbf{first} one is

\begin{equation}
P_{ij}\equiv \left[ \delta ,d\right] g_{ij}(\mathbf{u,v)}\equiv \delta
dg_{ij}(\mathbf{u,v})-d\delta g(\mathbf{u,v})\equiv \frac{\partial g_{ij}}{%
\partial u}\left[ \delta ,d\right] u+(u\leftrightarrow v)\text{ \ \ \ \ \ ,}
\tag{13}
\end{equation}
where $(u\leftrightarrow v)$ means the same expression with interchanged $u$
and $v$. As for the \textbf{second }expression, it shall be presented in the
following compact and symmetric form

\begin{equation}
\left( \delta d^{2}-d\delta ^{2}\right) g_{ij}(\mathbf{u,v})\equiv
Q_{ij}=\delta \widetilde{Q}_{ij}(\mathbf{du,dv)-d\widetilde{Q}}_{ij}(\mathbf{%
\delta u,\delta v)}\text{ \ \ \ \ \ ,}  \tag{14}
\end{equation}
where $\widetilde{Q}_{ij}(\mathbf{du,dv)}$ is the quadratic form in respect
to $\mathbf{du}$ and $\ \mathbf{dv}$ from eq.(7)

\begin{equation}
\widetilde{Q}_{ij}(\mathbf{du,dv)\equiv }d^{2}g_{ij}(u,v)=\frac{\partial
^{2}g_{ij}}{\partial u^{2}}(\mathbf{du)}^{2}+2\frac{\partial g_{ij}}{%
\partial u}\mathbf{.}\frac{\partial g_{ij}}{\partial v}\mathbf{dudv+}\frac{%
\partial ^{2}g_{ij}}{\partial v^{2}}(\mathbf{dv)}^{2}\text{ \ \ \ \ ,}
\tag{15}
\end{equation}
$\mathbf{\widetilde{Q}}_{ij}(\mathbf{\delta u,\delta v)}$ is the same form,
but with $\mathbf{\delta u}$ and $\mathbf{\delta v}$.

By means of the last three expressions, expression (12)\ for the variation
of the action can be represented in the following form

\begin{equation}
\delta S\equiv \int \left[ \frac{\delta L^{(0)}}{\delta g_{ij}}\delta
g_{ij}+A_{ij}+\delta B_{ij}+dC_{ij}\right] d^{n}x\equiv 0\text{ \ \ \ \ \ \
\ \ ,}  \tag{16}
\end{equation}
where
\begin{equation}
\frac{\delta L^{(0)}}{\delta g_{ij}}\equiv \frac{\partial L}{\partial g_{ij}}%
-d\left[ \frac{\partial L}{\partial (dg_{ij})}\right] +\delta d\left[ \frac{%
\partial L}{\partial (d^{2}g_{ij})}\right] \text{ \ \ \ \ \ \ \ \ .}
\tag{17}
\end{equation}
This expression will represent the equation of motion for $\ u$, provided
also that the variations at the endpoints vanish and $\delta B_{ij}=0$. $%
A_{ij\text{ }}$is a term, which appears in (16) due to the assumption $%
\delta \neq d$ and will disappear when $\delta =d$

\begin{equation*}
A_{ij}(\mathbf{u,v)\equiv -\delta }\left[ \frac{\partial L}{\partial
(d^{2}g_{ij})}\right] \widetilde{Q}_{ij}(\mathbf{du,dv})+\mathbf{d}\left[
\frac{\partial L}{\partial (d^{2}g_{ij})}\right] \widetilde{Q}_{ij}(\mathbf{%
\delta u,\delta v})+
\end{equation*}

\begin{equation}
+\left( \frac{\partial g_{ij}}{\partial v}\frac{\partial ^{2}L}{\partial
u\partial (dg_{ij})}-\frac{\partial g_{ij}}{\partial u}\frac{\partial ^{2}L}{%
\partial v\partial (dg_{ij})}\right) (du\delta v-\delta udv)\text{ \ \ \ \ \
\ \ .}  \tag{18}
\end{equation}
$B_{ij}$ is the expression

\begin{equation}
B_{ij}\equiv \frac{\partial L}{\partial (dg_{ij})}dg_{ij}-d(\frac{\partial L%
}{\partial (d^{2}g_{ij})})\delta g_{ij}+\frac{\partial L}{\partial
(d^{2}g_{ij})}\widetilde{Q}_{ij}(du,dv)\text{ \ \ \ \ \ \ \ ,}  \tag{19}
\end{equation}
and $C_{ij}$ is equal to

\bigskip
\begin{equation}
C_{ij}\equiv \frac{\partial L}{\partial (d^{2}g_{ij})}\delta ^{2}g_{ij}-%
\frac{\partial L}{\partial (d^{2}g_{ij})}\widetilde{Q}_{ij}(\mathbf{du,dv})=2%
\frac{\partial L}{\partial (d^{2}g_{ij})}K_{ij}\delta u\delta v\text{ \ \ \
\ \ \ \ \ \ .}  \tag{20}
\end{equation}
In (20) $K_{ij\text{ }}$ is given by

\begin{equation}
K_{ij}=\frac{\partial ^{2}g_{ij}}{\partial u\partial v}-\frac{\partial g_{ij}%
}{\partial u}\frac{\partial g_{ij}}{\partial v}\text{ \ \ \ \ \ \ .}
\tag{21}
\end{equation}
Note that \ if the variations at the initial and final endpoints of the
chosen curves disappear and also $\delta =d,$ then the equation $%
K_{ij}\equiv 0$ (taking into account second variation of the metric tensor $%
\delta ^{2}g_{ij}$) appears as a necessary and sufficient condition for the
fulfillment of the equations of motion (17).

\section{\protect\bigskip First Variation of the Lagrangian as a Third-Rank
Polynomial.}

From the representation (16) of the variated Lagrangian, let us express all
variations and differentiations in terms of the vector fields $u$ and $v$.
The following notations are introduced

\begin{equation*}
X_{1}^{i}\equiv \delta u^{i}\text{ \ \ \ \ \ \ \ \ \ \ }X_{2}^{i}\equiv
\delta v^{i}\text{ \ \ \ \ \ \ \ \ }Z_{1}^{i}\equiv \delta du^{i}\text{ \ \
\ \ \ \ \ \ \ \ \ \ \ }Z_{2}^{i}\equiv \delta dv^{i}
\end{equation*}
\begin{equation}
\text{ \ \ \ \ \ \ \ \ \ \ \ \ }Y_{1}^{i}\equiv du^{i}\text{ \ \ \ \ \ \ \ \
\ \ }Y_{2}^{i}\equiv dv^{i}\text{\ \ \ \ \ \ \ }T_{1}^{i}\equiv d\delta u^{i}%
\text{\ \ \ \ \ \ \ \ \ \ \ \ \ \ \ \ \ \ }T_{2}^{i}\equiv d\delta v^{i}%
\text{\ \ \ \ \ \ \ \ \ .\ }  \tag{22}
\end{equation}
After some transformations, an expression for the \textbf{first variation}
of the Lagrangian will be obtained:

\bigskip
\begin{equation*}
\delta L\equiv \left\{ P_{1}^{u}X_{1}+\widetilde{P}%
_{1}^{v}X_{2}+Q_{1}^{u}Z_{1}+\widetilde{Q}_{1}^{v}Z_{2}\right\} +
\end{equation*}
\begin{equation*}
+\left\{ P_{2}^{u}X_{1}Y_{1}+P_{3}^{u}Y_{1}X_{2}+\widetilde{P}%
_{2}^{v}X_{2}Y_{2}+\widetilde{P}%
_{3}^{v}Y_{2}X_{1}+Q_{2}T_{1}X_{2}+Q_{2}T_{2}X_{1}\right\} +
\end{equation*}

\begin{equation*}
+\left\{ P_{4}^{u}X_{1}X_{2}Y_{1}+\widetilde{P}%
_{4}^{v}X_{1}X_{2}Y_{2}+P_{5}^{u}X_{1}Y_{1}Y_{2}+\widetilde{P}%
_{5}^{v}X_{2}Y_{1}Y_{2}\right\} +
\end{equation*}
\bigskip
\begin{equation}
+\left\{ P_{6}^{u}X_{1}Y_{1}^{2}+\widetilde{P}%
_{7}^{v}X_{1}Y_{2}^{2}+P_{7}^{u}X_{2}Y_{1}^{2}+\widetilde{P}%
_{6}^{v}X_{2}Y_{2}^{2}\right\} \equiv 0\text{ \ \ \ \ \ \ \ .}  \tag{23}
\end{equation}
If we assume (just from a general consideration and not from a concrete
example) that the coefficient functions $P_{1}^{u},\widetilde{P}%
_{1}^{v},......,.P_{7}^{u},\widetilde{P}_{7}^{v},Q_{1},Q_{2}$ are
independent of the variables $X_{1},X_{2}........T_{1},T_{2}$ and depend
only on $u$ and $v$, then (23) will represent a \textbf{non-homogeneous
polynomial of third degree} (the highest degree in the polynomial) and it is
\textbf{linea}r in respect to the variables $X_{1},X_{2},Z_{1},Z_{2}$,
\textbf{quadratic} in respect to $X_{1},X_{2},Y_{1},Y_{2},T_{1},T_{2}$ and
\textbf{cubic} \textit{only }in respect to $X_{1},X_{2},Y_{1},Y_{2}$. Note
that the polynomial structure is rather specific, since $Z_{1},Z_{2}$ enter
only the \textbf{linear terms} and $T_{1},T_{2}$ only the \textbf{quadratic
terms (}and in combination only with $X_{1},X_{2}$ and not $Y_{1},Y_{2}$).
In (23) $P_{1}.......P_{7},Q_{1},Q_{2}$ are functions of $u$ and $v$ , which
are of the following form

\begin{equation}
P_{1}^{u}\equiv \frac{\partial L}{\partial g_{ij}}\frac{\partial g_{ij}}{%
\partial u}\text{ };\text{\ \ \ }P_{2}^{u}\equiv \frac{\partial L}{\partial
(dg_{ij})}\frac{\partial ^{2}g_{ij}}{\partial u^{2}}\text{\ \ \ \ \ \ \ \ \
\ \ \ ,}  \tag{24}
\end{equation}
\begin{equation}
P_{3}^{u}\equiv \frac{\partial }{\partial v}\left[ \frac{\partial L}{%
\partial (dg_{ij})}\frac{\partial g_{ij}}{\partial u}\right] -\frac{\partial
}{\partial u}\left[ \frac{\partial L}{\partial (dg_{ij})}\frac{\partial
g_{ij}}{\partial v}\right] +\frac{\partial L}{\partial (dg_{ij})}\frac{%
\partial ^{2}g_{ij}}{\partial u\partial v}\text{ \ \ \ \ ,}  \tag{25}
\end{equation}
\begin{eqnarray}
P_{4}^{u} &\equiv &2\frac{\partial L}{\partial (d^{2}g_{ij})}\frac{\partial
K_{ij}}{\partial u}\text{ \ \ \ \ \ }P_{5}^{u}\equiv -2\frac{\partial ^{2}L}{%
\partial u\partial (d^{2}g_{ij})}\text{\ \ \ \ \ \ \ \ \ \ \ \ \ ,}
\TCItag{26} \\
\text{\ }P_{6}^{u} &\equiv &-\frac{\partial ^{2}L}{\partial u\partial
\{d^{2}g_{ij})}\frac{\partial ^{2}g_{ij}}{\partial u^{2}}\text{ \ \ \ \ \ \
\ \ \ }P_{7}^{u}\equiv -\frac{\partial ^{2}L}{\partial v\partial
(d^{2}g_{ij})}\frac{\partial ^{2}g_{ij}}{\partial u^{2}}\text{ \ \ \ \ \ \ \
.}  \notag
\end{eqnarray}
$\widetilde{P}_{1}^{v}.....\widetilde{P}_{7}^{v}$ are the same expressions,
but with $u$ and $v$ interchanged. It should be noted, however, that
depending on the concrete Lagrangian, powers in $du,dv,\delta u,\delta v$
may come also from the coefficient functions, therefore changing the highest
degree of the polynomial. Further the problem shall be formulated in terms
of a third-rank polynomial, keeping in mind that the degree of the
polynomial does not change the formulation, but only the technical methods
for solving the algebraic equation.

\section{\protect\bigskip Third--Rank Polynomials - Formulation of the
Problem from an Algebro-Geometric Point of View.}

\bigskip Let us formulate the investigated problem from the point of view of
algebraic geometry, using the well-known approaches and terminology in
[11-13]. For a clear and illustrated with many examples exposition of the
subject, one may see also [14].

Let the defined in (22) set of variables $%
(X_{1}^{i},X_{2}^{i},Y_{1}^{i},Y_{2}^{i},Z_{1}^{i},Z_{2}^{i},T_{1}^{i},T_{2}^{i})
$ $(i=1,2...,n)$ $\ $belong

to the \textit{algebraic variety} $\overline{X}$\bigskip $\subset A^{n}(k)$,
where $A^{n}(k)$ is the $n-$\textit{dimensional affine space}, defined over
the field $\ k=k[X_{1},X_{2},Y_{1},Y_{2},Z_{1},Z_{2},T_{1},T_{2}]$ of the
functions in $8-$ variables. The coefficient functions $P_{1}^{u},\widetilde{%
P}_{1}^{v},P_{2}^{u},\widetilde{P}_{2}^{v},.......P_{7}^{u},\widetilde{P}%
_{7}^{v},Q_{1},Q_{2}$ \ are defined not on the same field $k$, but on the
manifold $M$. In fact, $A^{n}(k)$ \ is the \ cartesian product of $n-$tuples
of $k$. Since all the the components of the metric tensor and of the vector
fields are also defined on $M\supset (x_{1}^{0},x_{2}^{0},.......x_{n}^{0}),$
for each point $(x_{1}^{0},x_{2}^{0},.......x_{n}^{0})$ a mapping $\varphi :$
$M\rightarrow \overline{X}$ between the elements of the \textit{manifold }%
and the elements of the \textit{algebraic variety }is also defined

\begin{equation}
\varphi (x_{1}^{0},x_{2}^{0},..x_{n}^{0})=(X_{1}^{i}(x_{1}^{0},...x_{n}^{0}),%
\text{ }%
X_{2}^{i}(x_{1}^{0},...x_{n}^{0}),....,T_{1}^{i}(x_{1}^{0},...,x_{n}^{0}),T_{2}^{i}(x_{1}^{0},..,x_{n}^{0})%
\text{ \ \ \ \ \ .}  \tag{27}
\end{equation}

Now let $%
R[X_{1}^{i},X_{2}^{i},Y_{1}^{i},Y_{2}^{i},Z_{1}^{i},Z_{2}^{i},T_{1}^{i},T_{2}^{i}]
$ denotes the ring of all polynomials $f_{1},f_{2}.....,f_{n}...,$ defined
on the points $\overline{X}$ of the \textit{algebraic variety} $\subset
A^{n}(k)$ and belonging to the ideal $V(J,k),$ such as

\begin{equation}
V(J,k)=\left\{ \overline{X}\bigskip \subset A^{n}(k)\text{ }:\text{ }(f_{1}(%
\overline{X}),\text{ }f.(\overline{X}),....,f_{n}(.\overline{X}%
)...)=0\right\} \text{ \ \ \ \ \ \ \ .}  \tag{28}
\end{equation}
Therefore, from (27) and (28) it is easily seen that the following sequence
of mappings is defined

\begin{equation}
M\rightarrow \overline{X}\longrightarrow V(J,k)\text{ \ \ \ \ \ \ \ .}
\tag{29}
\end{equation}
\bigskip The considered in this paper problem can be defined in the
following way:

\begin{proposition}
.The variational problem $\delta L=0$ is equivalent to \textbf{finding all
the elements} $%
X_{1}^{i},X_{2}^{i},Y_{1}^{i},Y_{2}^{i},Z_{1}^{i},Z_{2}^{i},T_{1}^{i},T_{2}^{i}
$ of the algebraic variety $\overline{X}$, which satisfy \textbf{an
algebraic equation} $f(\overline{X})\equiv 0$, defined on the elements of
the variety and with a finite number of coefficient functions $P_{1}^{u},%
\widetilde{P}_{1}^{v},P_{2}^{u},\widetilde{P}_{2}^{v},.......P_{7}^{u},%
\widetilde{P}_{7}^{v},Q_{1},Q_{2}$ - functions of the metric tensor and the
two chozen vector fields. The algebraic equation belongs to the ideal $%
I=\left( P_{1}^{u},\widetilde{P}_{1}^{v},P_{2}^{u},\widetilde{P}%
_{2}^{v},.......P_{7}^{u},\widetilde{P}_{7}^{v},Q_{1},Q_{2}\right) \subset
R[X_{1}^{i},X_{2}^{i},Y_{1}^{i},Y_{2}^{i},Z_{1}^{i},Z_{2}^{i},T_{1}^{i},T_{2}^{i}],
$where $R$ is the ring of all the polynomials, defined on $\overline{X}$.
\end{proposition}

\bigskip If \ found and considered as functions, defined on the manifold $M,$
the elements of the algebraic variety $\overline{X}$ are no longer
independent, but will represent a system of \textbf{partial differential
equations} in respect to $du$ (or $\delta u$, for example). Further, if $du$
is known, then again the new system of partial differential equations (this
time in respect to $u$) will give an expression for $u$ (if $v$ is assumed
to be known), or it will give a relation between the two vector fields (if $%
u $ and $v$ are not known). Furthermore, the obtained relation between the
vector fields from the variational principle $\delta L=0$ might be used in
the determination of the equation of motion for $u.$ In an algebraic
language, the simultaneous investigation of the variation equation $\delta
L=0$ and the equation of motion means that the \textit{intersection
''varieties''} of the \textit{two algebraic surfaces} (defined by the
corresponding algebraic equations) should be found.

\begin{acknowledgement}
The author is grateful to Dr.S.Manoff from the Institute for Nuclear
Research and Nuclear Energy of the Bulgarian Academy of Sciences for some
interesting discussions, as well as to Dr.L.Alexandrov and \ Prof.N.A.
Chernikov (Laboratory for Theoretical Physics, JINR, Dubna) \ for their
interest towards this work and encouragement.
\end{acknowledgement}

\end{document}